\documentclass[pre,aps,twocolumn,reprint]{revtex4-2}

\usepackage{graphicx}
\usepackage{amssymb,latexsym,amsthm,amsmath}    
\usepackage{mathtools}
\usepackage{float} 
\usepackage[draft]{changes}
\usepackage{hyperref}
\usepackage{color}
\usepackage{hhline}
\usepackage{colortbl}
\usepackage{ulem} 
\usepackage[dvipsnames]{xcolor}

\newcommand{\mb}[1]{\mbox{\bfseries \itshape #1}}

\begin{document}

\title{From chaotic itinerancy to intermittent synchronization in complex 
networks}

\author{I. Leyva \& Irene Sendi\~na-Nadal} 

\affiliation{Complex Systems Group \& GISC, Universidad Rey Juan Carlos, 28933 M\'ostoles, Spain}

\affiliation{Center for Biomedical Technology, Universidad Polit\'ecnica de Madrid, Pozuelo de Alarc\'on, 28223 Madrid, Spain}

\author{Christophe Letellier}

\affiliation{Rouen Normandy University --- CORIA, Avenue de l'Universit\'e,
76800 Saint-Etienne du Rouvray, France}

\author{R. Sevilla-Escoboza}

\affiliation{Centro Universitario de los Lagos, Universidad de Guadalajara, Lagos de Moreno, Jalisco, Mexico}

\author{V. P. Vera-Ávila}

\affiliation{Instituto Polit\'ecnico Nacional, Unidad Profesional Interdisciplinaria de Ingenier\'ia Campus Guanajuato, Mexico}

\begin{abstract}
Although synchronization has been extensively studied, important processes underlying its emergence have remained hidden by the use of global order 
parameters. Here, we uncover how the route unfolds through a sequential 
transition between two well-known but previously unconnected phenomena: 
\textit{chaotic itinerancy} (CI) and \textit{intermittent synchronization} 
(IS).  Using a new symbolic dynamics, we show that CI emerges as a collective yet unsynchronized exploration of different domains of the high-dimensional attractor, whose dimension is reduced as the coupling increases, ultimately collapsing back into the reference chaotic attractor of an individual unit. At this stage, the IS can emerge as irregular alternations between synchronous and asynchronous phases. 
The two phenomena are therefore mutually exclusive, each dominating a distinct 
coupling interval and governed by different mechanisms. 
Network structural heterogeneity enhances itinerant behavior since access to different domains of the attractor depends on the 
nodes' topological roles.  The CI--IS crossover occurs within a consistent 
coupling interval across models and topologies. Experiments on electronic 
oscillator networks confirm this two-step process, establishing a unified framework for the route to synchronization in complex systems. 
\end{abstract}

\maketitle


\section{Introduction}
\label{intro}

The study of collective dynamics in complex networks has advanced our understanding of how local interactions give rise to emergent behaviors such as synchronization, criticality, and pattern formation. Among these, synchronization has been the main focus, particularly in strongly coupled systems where coherent global states emerge. This emphasis, however, has often obscured the dynamical richness of weakly coupled systems, which remain largely unexplored despite their relevance to many natural and artificial systems.  The brain is the most prominent example, where the role of criticality has long been 
recognized,\cite{Beggs2003,Shew2013,Munoz2018,Calvo2024} but similar self-organized critical behavior has also been identified in other complex systems—from gene networks and ecosystems to artificial neural ensembles \cite{Kauffman1993,Sole2006,Bertschinger2004}. Operating near criticality allows such systems to balance robustness and sensitivity, maximizing adaptability and computational capacity \cite{Deco2021}. 

When functioning away from global coherence, these systems display a rich 
repertoire of regimes and critical dynamics, including the two phenomena central to this work: \emph{chaotic itinerancy} (CI) \cite{Tsuda1992,Kay2003} and \emph{intermittent synchronization} (IS) \cite{Vera2020,choudhary2017}. 
IS, observed at the edge of global synchronization, involves irregular alternations between coherent and unsynchronized phases \cite{Platt1993,Heagy1994}. It corresponds to the well-known \textit{on–off} intermittency \cite{Vera2020,choudhary2017} and has been reported in neural activity under both healthy and epileptogenic conditions \cite{Breakspear2017,Bor23}. IS thus provides a mechanism by which complex systems self-tune to operate near criticality, though its dependence on the underlying network structure remains poorly understood.

In contrast, CI refers to the trajectory within a 
high-dimensional attractor wandering through a sequence of 
metastable regimes (also called \textit{attractor 
ruins}) connected \cite{Kan03,Tsuda2015,Koch2024} First introduced by Kaneko 
in nonequilibrium neural dynamics \cite{Kaneko1990}, CI has been observed in diverse contexts, including sensory processing in the olfactory cortex \cite{Kay2003}, perceptual switching in EEG activity \cite{Freeman2003}, laser dynamics \cite{Otsuka1991,Iwami2022}, and adaptive robotic behavior  \cite{Aucouturier2008,Steingrube2010,Park2017,Recanatesi2022}. 
In neural systems, metastable itinerant dynamics are thought to underlie 
cognitive processes such as working memory, decision making, and behavioral 
flexibility.\cite{Sasaki2007,Morrison2022,Roberts2019,Ashwin2024,Hancock2025,Rossi2025} 

These features have also made CI relevant to machine learning and computation. Reservoir computing, for instance, exploits high-dimensional transients for memory and pattern classification, and recent work has demonstrated controlled CI in such architectures \cite{Inoue2020,Kong2024,OHagan2025,Kab25}. This 
supports the idea that transient, metastable dynamics
are essential for flexible and context-dependent computation in both biological and artificial systems \cite{Koch2024,Mattera2025,Rossi2025,Hancock2025}.

Despite extensive studies of CI in globally coupled or random networks \cite{Kaneko1990,Tsuda1992,Uchiyama2006,Kong2024,Kab25}, its relationship with network topology remains poorly understood. Yet real-world systems exhibiting CI often feature modular, hierarchical, or heterogeneous architectures \cite{Torres2008,Koch2024,Khatun2024}. 

The main goal of this work is to address this gap. Specifically, we investigate how network topology influences the emergence, organization, and persistence of chaotic itinerancy in weakly coupled oscillator systems, and clarify its connection with IS, as the two phenomena are often confounded \cite{Sameshima2003}. We show that CI and IS are successive, mutually exclusive stages along the route to synchronization, each associated with distinct critical behaviors. These properties are robust across topologies, network sizes, node dynamics, and are further confirmed experimentally in electronic oscillator networks.

To track CI, we introduce a new symbolic dynamics approach that encodes 
nodal trajectories into symbolic sequences, capturing transitions 
with greater sensitivity than standard time-averaged measures. This framework 
allows the node-level analysis of visiting sequences and reveals how 
topological roles modulate access to different nodal regimes. It thus provides 
new insight into how the network topology shapes dynamics in complex dynamical 
systems ---bridging a gap between network theory and nonlinear dynamics, may 
prove useful for understanding the computational and decision processes driven by chaotic itinerancy.


\section{Chaotic itinerancy and intermittent synchronization}
\label{sec:backgroung}

The characterization of the route to synchronization in coupled oscillator 
networks is often described in terms of global, time-averaged metrics, such as the mean synchronization error or an order parameter that summarizes 
collective coherence. However, such global indicators often hide the 
underlying, time-dependent phenomena occurring at the nodal level. In 
particular, they hide two critical yet interconnected processes: chaotic 
itinerancy (CI) and intermittent synchronization (IS). CI corresponds to 
irregular transitions at the node level among metastable regimes, whereas IS 
manifests as intermittent alternations between synchronous and asynchronous 
phases at the network scale. Both arise in weak-to-intermediate coupling 
values, revealing the multiscale organization of dynamics that bridges nodal
and network levels of the synchronization process.

To explore these regimes, we consider a network of 
$N$ diffusively coupled identical chaotic oscillators governed by
\begin{equation}
   \label{eq:dynet}
  \dot{\mathbf{\mb{x}}}_i 
	= \mathbf{f}(\mathbf{\mb{x}}_i) 
	- d \sum_{j=1}^N{L}_{ij}\,\mathbf{h}(\mb{x}_j)
\end{equation}
where $\mathbf{\mb{x}}_i \in \mathbb{R}^m$ is the state vector of the $i$th 
node, $\mathbf{f}$ and $\mathbf{h}$ are $\mathbb{R}^m\rightarrow\mathbb{R}^m$ 
functions 
governing
the intrinsic dynamics of an isolated unit and the 
coupling between them respectively; $d$ is the coupling strength, $L_{ij}$ is 
the Laplacian matrix encoding the network topology, with ${L}_{ij}=-1$ if 
nodes $i$ and $j$ are connected (${L}_{ij}=0$ otherwise), and ${L}_{ii}=k_i$ 
is the degree of the $i$th node. This framework captures a broad class of 
diffusive interactions and allows us to systematically track the underlying 
multiscale organization of the synchronization process. The network is a single 
dynamical system whose state space is $\mathbb{R}^{Nm}$, 
and producing an attractor 
${\cal A} (d) \subset \mathbb{R}^{Nm}$ depends on
 the coupling value $d$, the other parameter values being kept constant. 
Indeed,
each node $i$ is associated with a $m$-dimensional sub-state space 
$\mathbb{R}^m_i$ in which there is an attractor
${\cal A}_i(d) \subset \mathbb{R}_i^m$. When uncoupled $d = 0$, the 
subattractors ${\cal A}_i (0)$
are identical to ${\cal A}_{\rm r}$ that will be in the next Sections named as 
the \emph{reference attractor}.


\begin{figure}[ht!]
  \centering
  \includegraphics[width=\linewidth]{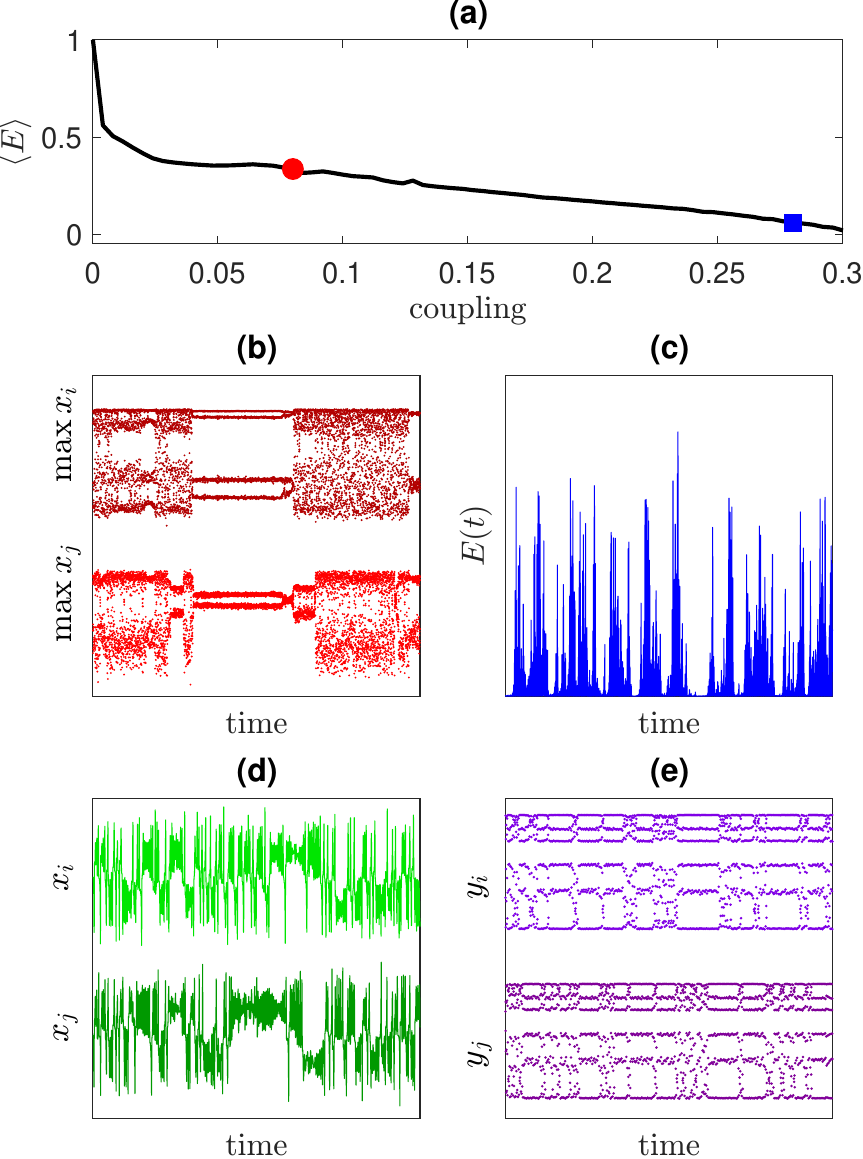}
  \caption{Examples of CI and IS. (a) Mean synchronization error 
$\langle E \rangle$ of a network of R\"ossler units, highlighting two points 
along the route to synchronization: the red dot marking a coupling regime 
where chaotic itinerancy (CI) takes place, and at higher coupling, the blue square pointing out a regime of intermittent synchronization (IS). (b) Evidence of the CI marked as a red point in (a) ($d=0.08$), showing the maxima of the $x$ variable of two arbitrary nodes in the network  (shifted for visualization).  (c) Evidence of the IS marked as a blue square in (a) ($d=0.28$) using the instantaneous synchronization error $E(t)$. (d) Example of CI showing the $x$ variable (shifted for clarity) of two arbitrary Lorenz nodes from a larger clique configuration. (e) Globally coupled $N$=100 logistic maps as an example of a discrete dynamical system showing CI. Models used in Eq. \eqref{eq:dynet}: (a) 
and (b) $N=100$ R\"ossler nodes coupled in a scale-free configuration with average node degree $\langle k\rangle=4$, with 
$\mathbf{f}(\mathbf{x}) = \left[-y-z,x+ a y,z(x-4)+2\right]$ and 
$\mathbf{h}(\mathbf{x})=[0,y,0]$, with $a=0.432$; (d) a clique graph of $N=5$ Lorenz systems with $\mathbf{f}(\mathbf{x}) = \left[-10(y-x),x(28-z)-y,xy-2z\right]$, $\mathbf{h}(\mathbf{x})=[0,y,0]$. Parameters used in (d) for globally coupled Logistic 
maps, $y_i(t+1)=f\left((1-d)y_i(t)+\frac{d}{N}\sum_{j=1}^N y_j(t)\right)$ with $f(x)=4(1-x^2)$ and $d=0.1$.}
  \label{fig:intro}
\end{figure}

Figure~\ref{fig:intro} illustrates the emergence of the two dynamical 
phenomena discussed above. Panel~(a) shows the expected gradual decrease of 
the synchronization error 
\begin{equation}
  \langle E\rangle = \lim_{T\to\infty} (1/T) \int_0^T E(t)\,dt \, , 
\end{equation}
with \(E(t)=\frac{2}{N(N-1)}\sum_{i<j}\|\mathbf{x}_i(t)-\mathbf{x}_j(t)\|^2\), 
for a network of \(N=100\) diffusively coupled R\"ossler oscillators. 

Along this synchronization process, two key phenomena are typically 
overlooked: at low coupling [red dot at $d = 0.08$, Fig.\ \ref{fig:intro}(b)], CI dominates, here represented by the trajectories of two arbitrary nodes (we plot the maxima of their two $x$-variables, vertically shifted for clarity). At higher couplings [blue square at $d = 0.28$ in Fig.\ \ref{fig:intro}(c)], IS becomes the dominant feature, where the instantaneous synchronization error 
$E(t)$ exhibits intervals of near-perfect synchrony interrupted by 
desynchronization bursts [Fig.\ \ref{fig:intro}(c)]. 

Further examples shown in Fig.\ \ref{fig:intro}(d) for a small complete 
graph of Lorenz oscillators and in Fig.\ \ref{fig:intro}(e) a network of 
globally coupled Logistic maps \cite{Kaneko1990}, confirm the generality of 
these behaviors across different models and network structures. 

These observations underscore the need for analytical tools that outperform 
ensemble-averaged measures and can capture the multiscale nature of these 
phenomena along the synchronization process. In the following subsections, we 
introduce two dedicated metrics: the CI and IS indexes. 

\subsection{Chaotic itinerancy index}
\label{sec:CI_index}


\begin{figure*}[t!]
  \centering
\includegraphics[width=\linewidth]{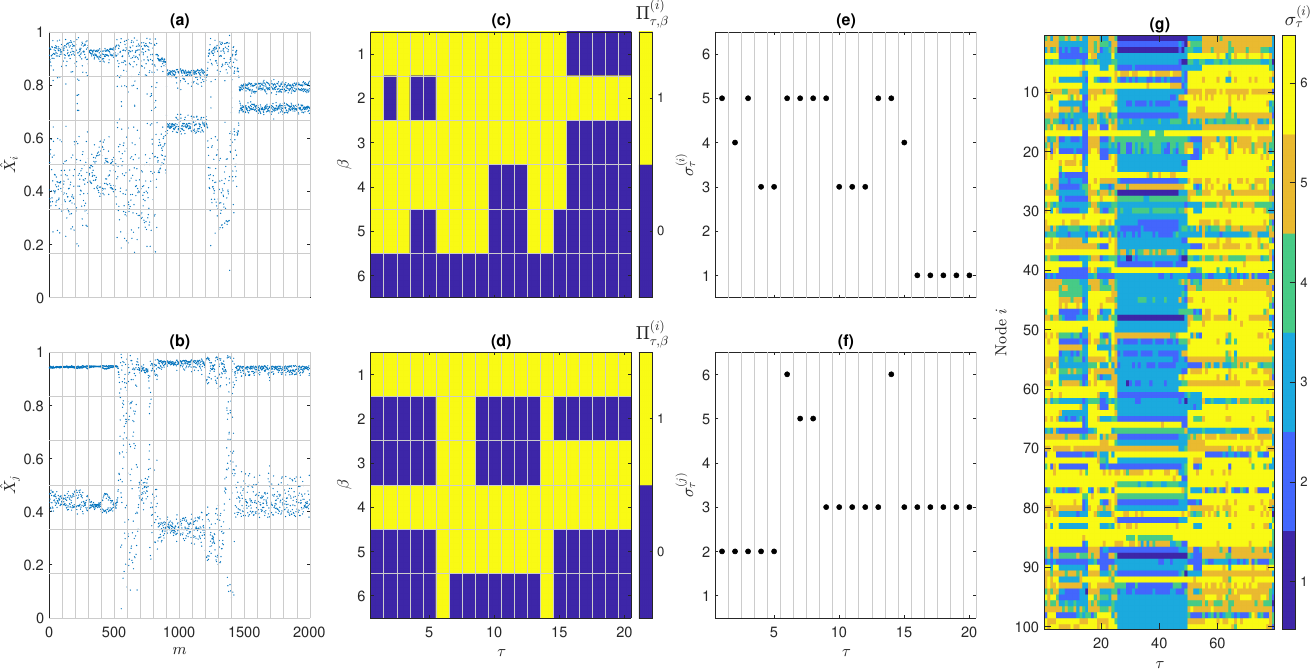}
  \caption{Symbolic encoding procedure used to characterize chaotic 
itinerancy, illustrated for two arbitrary nodes in a network of coupled 
R\"ossler oscillators.  
(a, b) Normalized sampled time series \( \hat{X}_i \) and \( \hat{X}_j \), 
corresponding to the scalar variables \( X_i(t) \) and \( X_j(t) \) of two 
arbitrary nodes, obtained here as the sequence of maxima of the R\"ossler 
variable \( x \). The data points are distributed over a grid (gray lines) composed of \( N_{\tau} \times N_{\beta} \) cells, where each time window \( \tau \) contains \( D \) consecutive samples. In this example, \( D = 100 \) and only 20 time windows (out of the total \( N_{\tau} \)) are shown.  
(c, d) Binary occupancy maps \( \Pi^{(i)}_{\tau,\beta} \) for the nodes shown 
in panels (a) and (b). Each cell \( (\tau, \beta) \) is marked as visited 
(\( \Pi^{(i)}_{\tau,\beta} = 1 \), yellow) if it contains more than \( \theta=2 \) points, and as empty (\( \Pi^{(i)}_{\tau,\beta} = 0 \), blue) otherwise.  
(e, f) Symbolic state \( \sigma^{(i)}_{\tau} \) for each time window, defined as the number of active cells visited by node \( i \) during window \( \tau \). In the example, full exploration of the spatial partition corresponds to \( \sigma^{(i)}_{\tau} = N_{\beta} = 6 \).  
(g) Symbolic raster plot \( \boldsymbol{\sigma} = \{ \sigma^{(i)}_{\tau} \} \) showing the symbolic evolution of all \( N \) nodes in the network. Example shown corresponds to a scale-free (SF) network of \( N = 100 \) coupled R\"ossler oscillators with coupling strength \( d = 0.08 \) and parameter 
\( a = 0.432 \). }
  \label{fig:symbol-setup}
\end{figure*}

Quantifying chaotic itinerancy (CI) in a network dynamics is challenging due 
to the high-dimensional wandering trajectories 
\cite{Tsuda1992,Kan03,Tsuda2015}. Previous attempts relied on Lyapunov 
exponents, recurrent plots or entropy rates to detect transitions between 
metastable regimes 
\cite{Timme2002,Freeman2003, Sameshima2003,Roberts2019,Khatun2024}. 
However, most of these approaches rely on global or averaged measures, thus 
overlooking the heterogeneous and node-dependent nature of transitions in 
networked systems. In order to overcome this limitation, we introduce a 
symbolic dynamics that captures CI directly at the nodal level, tracking the 
time sequence of transitions for each node individually. The goal is to reduce 
each oscillator's time series to a discrete set of symbols that reflect the 
sequence of regimes over time, enabling us to detect and quantify 
transitions between them. The process is exemplified in Fig. 
\ref{fig:symbol-setup} for two nodes of the $N=100$ R\"ossler oscillators 
network.  

Let \( {X}_i = \{ X_i(t_p) \}_{p=1}^P \) denote the scalar time series 
associated with the $i$th node, extracted from its state vector \( \mb{x}_i \) at discrete times $t_p$. The choice of variable and sampling strategy can be 
adapted to the system under study. In the Fig \ref{fig:symbol-setup} example, 
each node's time series is obtained by sampling the local maxima of the first variable, i.e. the \( X_i = x_i \) are the crossings of a Poincaré section defined by ${\cal P}_i \equiv \left\{ \displaystyle (x_i \in \mathbb{R} ~|~   \dot{x}_i =0, x_i >0  \right\} $. The series is normalized to the interval \([0,1]\) as \(\hat{X}_i=X_i/\max(X_i)\), ensuring comparability across nodes and realizations. 

Note that, for other dynamical systems or data, regular time sampling or local 
extrema of other variables can be used. The method does not rely on a specific 
type of sampling, as long as it captures relevant dynamical features. We will 
present examples in the next Sections. 

The normalized data are partitioned into \(N_\tau\) time windows of \(D\) 
consecutive samples each, and into \(N_\beta\) equally spaced amplitude bins. 
Each cell of this space-time grid is labeled by the indices \((\tau,\beta)\), 
where \(\tau \in \{1,\dots,N_\tau\}\) and \(\beta \in \{1,\dots,N_\beta\}\).  
In the example given in Fig.~\ref{fig:symbol-setup}(a,b), $N_{\beta}=6$, and 
only 20 time windows are shown from a much larger $N_\tau$. The window size 
$D$ has to be chosen such that there is a trade-off between capturing 
short-lived transitions with adequate time resolution and ensuring statistical 
reliability of the symbolic dynamics. Similarly, the number of amplitude bins 
$N_{\beta}$ is set to discretize the state-space exploration with sufficient 
granularity, while remaining robust to moderate changes; as later illustrated, the results are not critically sensitive to the specific choice of $D$ or $N_{\beta}$ (see Suppl. Material, Fig. S1).
 
A \((\tau,\beta)\) cell is considered visited if the number of points 
\(\pi_{\tau,\beta}^{(i)}\) falling inside exceeds a threshold \(\theta\):  
\begin{equation}
  \Pi_{\tau,\beta}^{(i)} =
  \begin{cases}
    1, & \text{if } \pi_{\tau,\beta}^{(i)} > \theta,\\
    0, & \text{otherwise.}
  \end{cases}
\label{eq:occupation}
\end{equation}
Figure~\ref{fig:symbol-setup}(c,d) illustrates this binary visitation grid for the 
two series in panels (a) and (b), respectively, where yellow encodes a visited
cell and blue a non-visited one.

The symbolic state of node \(i\) in window \(\tau\) is then defined as the 
number of visited cells 
\begin{equation}
  \sigma^{(i)}_{\tau} = \sum_{\beta=1}^{N_{\beta}} \Pi^{(i)}_{\tau, \beta}
  \label{eq:symbol}
\end{equation}
as shown in Figs.~\ref{fig:symbol-setup}(e,f). This symbolic 
dynamics \( \sigma^{(i)}_\tau\in \{1, \dots, N_{\beta}\} \) encodes the 
part of the attractor visited for that duration. Large values of \(\sigma_{\tau}^{(i)}\) 
correspond to a trajectory exploring wide domains of the attractor, whereas 
small values indicate confinement to near-periodic or low-developed
chaotic regimes.  

Applying this procedure to all nodes and time windows yields a 
\emph{symbolic raster plot} \( \boldsymbol{\sigma} = \{\sigma^{(i)}_\tau\} \) 
(Fig.~\ref{fig:symbol-setup}(g)), which captures the nodal diversity and time 
variability of the network's dynamics. 

The \emph{chaotic itinerancy index} is then defined as the average number 
of transitions between symbolic states per node and per window: 
\[
\Phi_\sigma = \frac{1}{N(N_\tau-1)} \sum_{i=1}^{N} 
\sum_{\tau=1}^{N_\tau-1}
\bigl[1-\delta_{\sigma_{\tau+1}^{(i)},\sigma_{\tau}^{(i)}}\bigr],
\tag{4}
\]
where 
\(\delta_{ab}\) is the Kronecker delta. High values of \(\Phi_\sigma\) 
indicate frequent transitions between metastable regimes ---strongly 
itinerant trajectory--- whereas small values reflect a single regime. 

\subsection{Intermittent Synchronization Index}

In contrast to chaotic itinerancy, the quantification of intermittent 
synchronization (IS) is a more established problem.  
On-off intermittence has been extensively studied in low-dimensional and small ensembles of coupled chaotic systems \cite{Platt1993,Gau96}, but the intermittence in complex networks has not been explored as much \cite{choudhary2017,Vera2020}.
Several approaches have been proposed to characterize this behavior, typically based on the statistics of laminar-phase durations, return maps, or burst amplitude distributions \cite{Heagy1994,Boccaletti2000}. 
Here, rather than introducing a new intermittency measure, we adopt a normalized definition that mirrors the structure of the CI index. 

While \(\Phi_\sigma\) quantifies CI by accounting for local transitions 
between metastable regimes, the equivalent \emph{intermittent 
synchronization index} \(\Phi_E\) should capture global alternations between 
coherent and incoherent network phases defining IS. 

It is computed from the instantaneous synchronization error \(E(t)\) introduced above. We binarize $E(t)$ using a threshold \(\theta_E\), defining 
$\varepsilon(t) = H\!\left(E(t) - \theta_E\right) $
where \(H(x)\) is the Heaviside step function \cite{Vera2020,Cho17}. The value \(\varepsilon(t)=0\) corresponds to (synchronous) phases, while \(\varepsilon(t)=1\) denotes desynchronized bursts. Typical practical choices for the threshold are small (e.g. \(\theta_E \sim 10^{-5}\)), but \(\theta_E\) should be set according to noise level.

The index is then defined as the normalized number of transitions between these two symbols:
\begin{equation}
  \Phi_E = \frac{1}{2}\frac{1}{T-1} 
	\sum_{t=1}^{T-1} | \varepsilon(t+1)-\varepsilon(t))|
  \label{eq:phi_Z}
\end{equation}
Large values of \(\Phi_E\) indicate frequent switching between 
coherent and non-coherent phases —signatures of on--off intermittency—
whereas \(\Phi_E=0\) corresponds either to complete incoherence or full 
synchronization. Thus, \(\Phi_E\) measures the instability of the laminar 
phases rather than the absolute level of coherence.
Both indices, \(\Phi_\sigma\) and \(\Phi_E\), provide complementary 
quantitative descriptions of the nodal level (CI) and network level (IS) 
dynamics that organize the route to synchronization in complex networks.


\section{Sequential Onset of CI and IS}
\label{seqonset}

 \begin{figure*}[ht!]
  \centering
	 \includegraphics[width=0.60\linewidth]{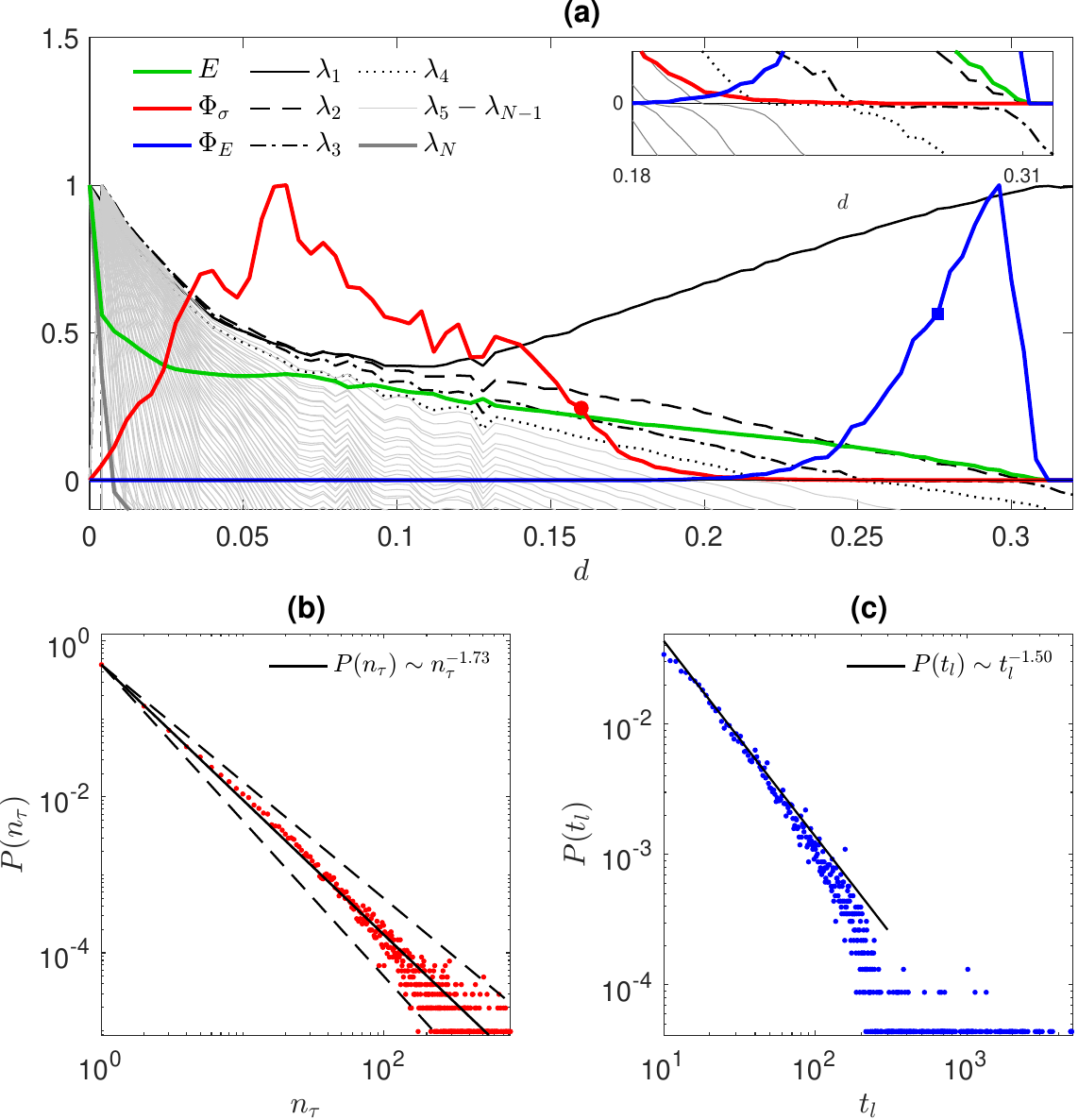} \\[-0.3cm]
  \caption{
Synchronization route and emergence of chaotic itinerancy (CI) and 
intermittent synchronization (IS) in a network of $N=100$ chaotic R\"ossler 
oscillators. (a) Global synchronization error $E$ (green), chaotic itinerancy 
index $\Phi_\sigma$ (red), and intermittent synchronization index $\Phi_E$ 
(blue) as a function of the coupling strength $d$. The $N$ largest Lyapunov 
exponents are also plotted with different colors of gray and line styles as 
indicated in the legend. The inset zooms the coupling interval where CI 
 vanishes and gives rise to IS until the system reaches global synchrony with 
$\lambda_1>0$, $\lambda_2=0$, $\lambda_3<0$.   
(b) Log-log probability distribution of the residence duration $n_\tau$ in the 
most developed chaotic symbolic state $\sigma=N_{\beta}$, at $d$ = 0.16, red 
 circle in (a), showing power-law behavior. The solid straight-line shows the 
best fit to a power law $P(n_\tau) \sim n_\tau^{-1.73}$, and the dashed 
straight-lines bounding the data correspond to exponents $2$ (upper) and 
 $1.5$ (bottom).
(c) Log-log probability distribution of the coherent phase durations $t_\ell$ 
during IS, at $d$ = 0.27 (blue square in (a)), following a $P(t_\ell) \sim t_\ell^{-3/2}$ scaling typical of on-off intermittency. Ensemble averages are computed over 30 independent realizations of initial conditions. Rest of parameters: $D = 200$, $N_\beta = 6$, $\theta = 3$. The network structure is an instance of a network with scale-free degree distribution and average degree 4.
\label{fig:joint_num_SFER}}
\end{figure*}

To investigate the onset and development of chaotic itinerancy and its 
relation to intermittent synchronization, we analyze the route to 
synchronization of an $N=100$ ensemble of identical Rossler oscillators in a 
heterogeneous undirected topology with $\langle k \rangle$=4. Such a topology
exhibits a sufficient degree of heterogeneity and is large enough to avoid the appearance of global periodic regimes, allowing us to focus on the genuinely chaotic itinerant dynamics that emerge during the route to synchronization. Statistical robustness is achieved by sampling a large number of different initial conditions for the oscillator states.

Figure~\ref{fig:joint_num_SFER}(a) reports the evolution of the CI index 
$\Phi_\sigma$ (red) and the IS index $\Phi_E$ (blue) as a function of the 
coupling strength $d$, together with the synchronization error $E$ (in green) plotted as a reference frame of the global coherence state. The key insight is that $\Phi_\sigma$ and $\Phi_E$ are activated consecutively and do not coexist: 
itinerancy is established immediately at low couplings and increases rapidly, 
reflecting the onset of a collective exploration of metastable regimes. The CI 
index reaches a maximum and then fades progressively as the oscillators 
converge onto the same symbolic state, corresponding to the reference 
chaotic attractor ${\cal A}_{\rm r}$.

Only after this contraction, IS does emerge as a signal that the final stage of the synchronization process has begun (around $d \sim 0.2$ in the example). During the IS regime, periods of near-synchrony are interrupted by global desynchronization bursts, in line with previous findings~\cite{Vera2020, Cho17}. Here, its role is reinterpreted as the natural continuation of the CI collapse, where all the nodes settled in the reference attractor ${\cal A}_{\rm r}$ but not in a synchronous way. The sequencing of these two dynamical regimes --- first CI, then IS--- is a  central contribution of this work.


A complementary view of the sequential onset of CI and IS comes from the 
Lyapunov spectrum of the full 3N‑dimensional system. We computed the spectrum 
using the standard Benettin algorithm \cite{Ben80} with Gram–Schmidt 
reorthonormalization on the tangent space of the network dynamics, following 
the procedure of Ref. \cite{Gut13}. The $N$ largest Lyapunov exponents are 
shown in Fig.~~\ref{fig:joint_num_SFER}(a) (shaded gray) overlaid on the CI 
and IS indicators. CI persists while the exponents $\lambda_N$ through 
$\lambda_4$ become negative sequentially, indicating the progressive loss of 
effective dimension that enabled wandering through different regimes. When $\lambda_4$ reaches zero [around $d\sim 0.22$ in the 
example; see inset in Fig.\ \ref{fig:joint_num_SFER}(a)], CI collapses, 
signaled by the vanishing of $\Phi_\sigma$. At that point, the synchronization 
process can begin, and the IS index $\Phi_\sigma$ increases as the system 
fluctuates near the synchronous manifold due to the only remaining marginal 
direction given by $\lambda_3\sim$0. The final collapse occurs when 
simultaneously $\lambda_3<0$ and $\lambda_2=0$, leaving ${\cal A} (d)$ 
being topologically equivalent to ${\cal A}_{\rm r}$. 

Equivalently, from the Lyapunov spectrum we can use the Kaplan-Yorke conjecture 
to calculate the Lyapunov dimension of the whole attractor as a function of the 
coupling, ${\cal D}_{\rm L}(d)$ \cite{Kap79}. From this point of view, CI takes 
place in the interval ${\cal D}_{\rm L}(0) = N{\cal D}_{\rm r} > 
{\cal D}_{\rm L} (d) > 4$, where ${\cal D}_{\rm r} = 2.03$  is the Lyapunov 
dimension of ${\cal A}_{\rm r}$ for the chosen parameters. Then, IS is observed 
in the sequent interval 
$4 \geqslant {\cal D}_{\rm L}(d) \geqslant {\cal D}_{\rm r} $. 
According to this R\"ossler network, 
chaotic itinerancy is only observed when the dynamics can be embedded in a 
space whose dimension is at least equal to 5. 


The correlation between the presence of near-zero Lyapunov exponents and the occurrence of chaotic itinerancy (CI) or intermittent synchronization (IS) has been discussed in previous works \cite{Kan94,Timme2002,Kan03,Tsuda2004,Tirabassi2025}, but a clear distinction between these two regimes and their specific correspondence with the Lyapunov spectrum in large-scale systems had not been clearly established.

To assess the presence of criticality in CI and IS, we analyze the 
distribution of residence durations within the symbolic state corresponding
to the reference attractor ${\cal A}_{\rm r}$,  encoded 
by the highest symbolic value $\sigma_i^\tau = N_\beta$. We therefore measure, for each node, the residence duration 
$n_{\tau}$ during which it remains in $N_\beta$  before switching 
onto another symbol. 
Figure~\ref{fig:joint_num_SFER}(b) shows that the resulting distribution
—averaged over all nodes and realizations— follows a power-law decay, 
$P(n_{\tau}) \sim n_{\tau}^{-1.7}$, for $d$ approaching 0.22, the coupling 
value at which CI ceases. The emergence of this scaling precisely at the point 
where the itinerant dynamics collapse supports the interpretation of a 
critical transition: as the nodes diminish their exploration of different regimes, its time organization becomes scale-free. 
Similar power-law residence-duration distributions have been reported in 
deterministic \cite{Sameshima2003} and stochastic \cite{Namikawa2025} models 
of chaotic itinerancy, where nodal trajectories wander among different 
regimes, although without explicitly identifying this final critical stage.

An analogous analysis was performed for the intermittent synchronization.  
Here, the relevant residence time corresponds to the duration of the 
synchronized phases, defined as the intervals during which the global 
synchronization error $E(t)$ remains below a small threshold before a 
desynchronization burst occurs. 
Figure~\ref{fig:joint_num_SFER}(c) shows that the distribution of these
synchronized phases durations \( t_\ell \) [measured at $d=0.27$, blue circle 
in Fig.~\ref{fig:joint_num_SFER}(a)] follows a power-law scaling, 
\( P(t_\ell) \sim t_l^{-1.5} \), consistent with the universal exponent 
reported for on--off intermittency \cite{Heagy1994,Platt1993,Heagy1994,Fuj83,Platt1993,Ham94b, Venkataramani95}. 

To further assess the consistency of the CI-IS sequence, we compared systems with different network architectures and node dynamics. 
Specifically, we analyzed networks with degree distributions ranging from complete graph to power-law, system sizes from $N=5$ to $N=500$ nodes, and distinct oscillator models by including the Lorenz system. Additionally, in this last case the analyzed data is not a Poincar\'e section, but a regular periodic sampling of the time series of the $x$ variable of the nodes, with $D=100$ and $N_\beta=2$, adapted to the different time scale and nature of the state space. 

\begin{figure}[ht]
  \centering
\includegraphics[width=\linewidth] {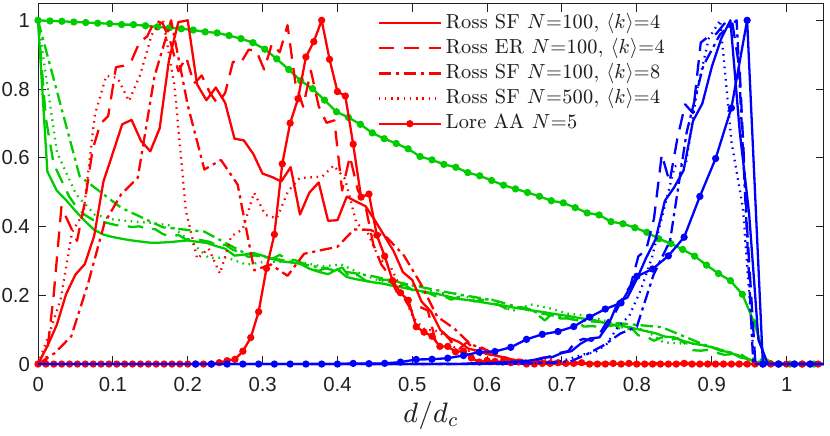} \\[-0.2cm]
  \caption{
Comparison of the chaotic itinerancy index $\Phi_\sigma$ (red) and intermittent synchronization index $\Phi_E$ (blue) for different combinations of network structures, sizes, and dynamical systems, distinguished by linestyles as indicated in the legend. The corresponding synchronization error $E$ (green) is also shown. In all cases, the coupling strength $d$ is normalized by the critical synchronization threshold $d_c$ of each network. Remarkably, despite the diversity of systems, the crossover point consistently occurs at $d/d_c \approx 2/3$. 
}
  \label{fig:robustness}
\end{figure}

Figure~\ref{fig:robustness} shows that when the coupling strength $d$ is 
rescaled by the corresponding critical synchronization threshold $d_c$ of each 
system, the crossover between the itinerant and intermittent phenomena, 
identified by the intersection of $\Phi_{\sigma}$ and $\Phi_E$, occurs at 
approximately $d/d_c \simeq 2/3$ in all cases. This robust scaling suggests a 
general organizational principle in the route to synchronization: 
independently of the nodal dynamics or network topology, the system 
transitions from locally itinerant to globally intermittent dynamics at the 
same point of the route to coherence. 

\section{CI nodal structure}
\begin{figure}[t!]
    \includegraphics[width=\linewidth]{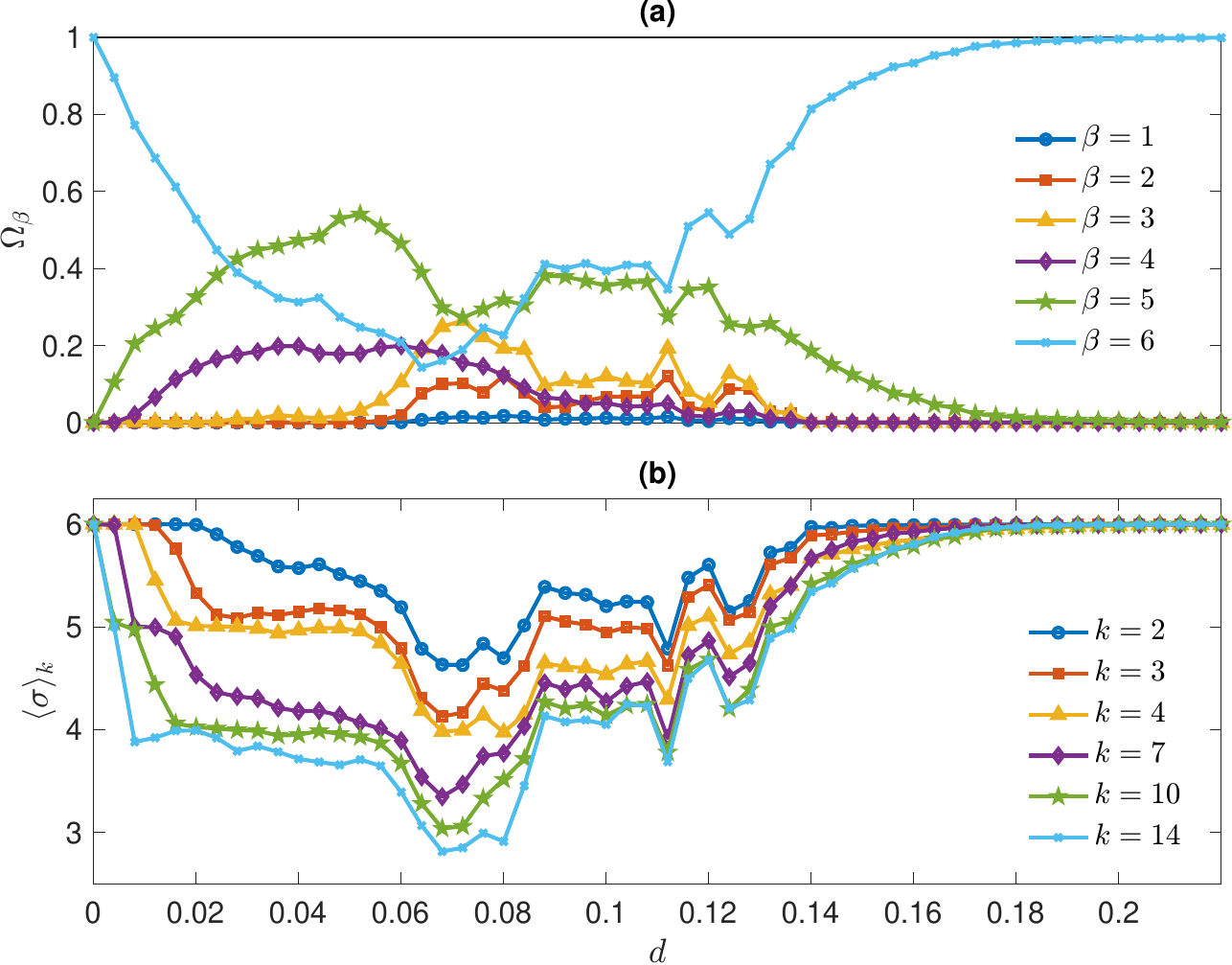}
  \caption{
Analysis of the symbolic state realization for the network used in 
Fig.~\ref{fig:joint_num_SFER}. 
(a) Realization probability of symbol $\Omega_\beta$ for each symbol 
$\beta = 1, \dots, N_\beta = 6$, and (b) average symbolic state per degree 
class $\langle \sigma \rangle_k$ as a function of coupling $d$. 
\label{fig:symbol_degree_evol}}
\end{figure}

The symbolic dynamics introduced in Sec. \ref{sec:CI_index} not only provides 
a quantitative characterization of chaotic itinerancy but also enables a 
detailed exploration of how the nodal structure of the network shapes 
chaotic itinerancy. By encoding each node's trajectory into $N_\beta$ 
symbols, we can directly assess how structural heterogeneity modulates the 
access and persistence of nodes within distinct regimes of the attractor
${\cal A}(d)$.

Figure~\ref{fig:symbol_degree_evol}(a) illustrates how the symbolic occupation 
statistics evolve across the coupling range where CI develops. The 
\emph{symbol probability} 
\begin{equation}
  \Omega_{\beta} = \frac{1}{N N_{\tau}} 
  \sum_{\tau=1}^{N_{\tau}} \sum_{i=1}^{N} 
  \delta_{\sigma^{(i)}_{\tau}, \beta},
\end{equation}
quantifies the relative frequency of each symbol 
$\beta \in \{1, \ldots, N_{\beta}\}$ across all nodes and time windows.
For $d = 0$, the highest symbolic state $\beta = N_\beta$  is the only one to be 
observed. As $d$ increases, additional symbols 
— representing more regular and less developed regimes — become progressively visited. Each symbol reaches its maximum realization at a specific coupling value [Fig.\
\ref{fig:symbol_degree_evol}(a)], revealing a \emph{hierarchical ordering} in 
which less chaotic regimes peak later along the route to synchronization. 
As $d$ approaches the critical value at which $\Phi_{\sigma}$ 
collapses ($d \sim 0.22$), the distribution reconverges toward $\Omega_{N_{\beta}} \approx 1$,  indicating that CI has effectively ceased.

Figure~\ref{fig:symbol_degree_evol}(b) reveals clear evidence of topologically induced dynamical differentiation~\cite{Tla19a,Let21b}. 
We define the \emph{degree-class symbolic average} as
\begin{equation}
  \langle \sigma \rangle_k = \frac{1}{|C_k| N_\tau} 
  \sum_{i \in C_k} \sum_{\tau=1}^{N_\tau} 
  \sigma^{(i)}_\tau,
\end{equation}
where $C_k$ is the set of nodes with degree $k$. The results indicate that the 
regimes accessible to each node depends strongly on its 
topological role. Highly connected nodes (hubs) explore a larger set of 
regimes and settle earlier into less chaotic ones, whereas 
low-degree nodes remain mostly confined in regimes close to the original 
chaotic regime. This degree-dependent differentiation produces a 
clear hierarchy in symbol exploration, linking topological heterogeneity to 
dynamical specialization during the collapse of itinerancy.

This degree-dependent differentiation may have broader implications for 
understanding how itinerance organize and interact in heterogeneous 
systems such as large-scale brain dynamics or in the design of reservoir 
computing architectures, where structured heterogeneity enhances computational 
flexibility and memory capacity.

\subsection*{Functional symbolic synchronization}
\begin{figure}[t!]
  \includegraphics[width=\linewidth]{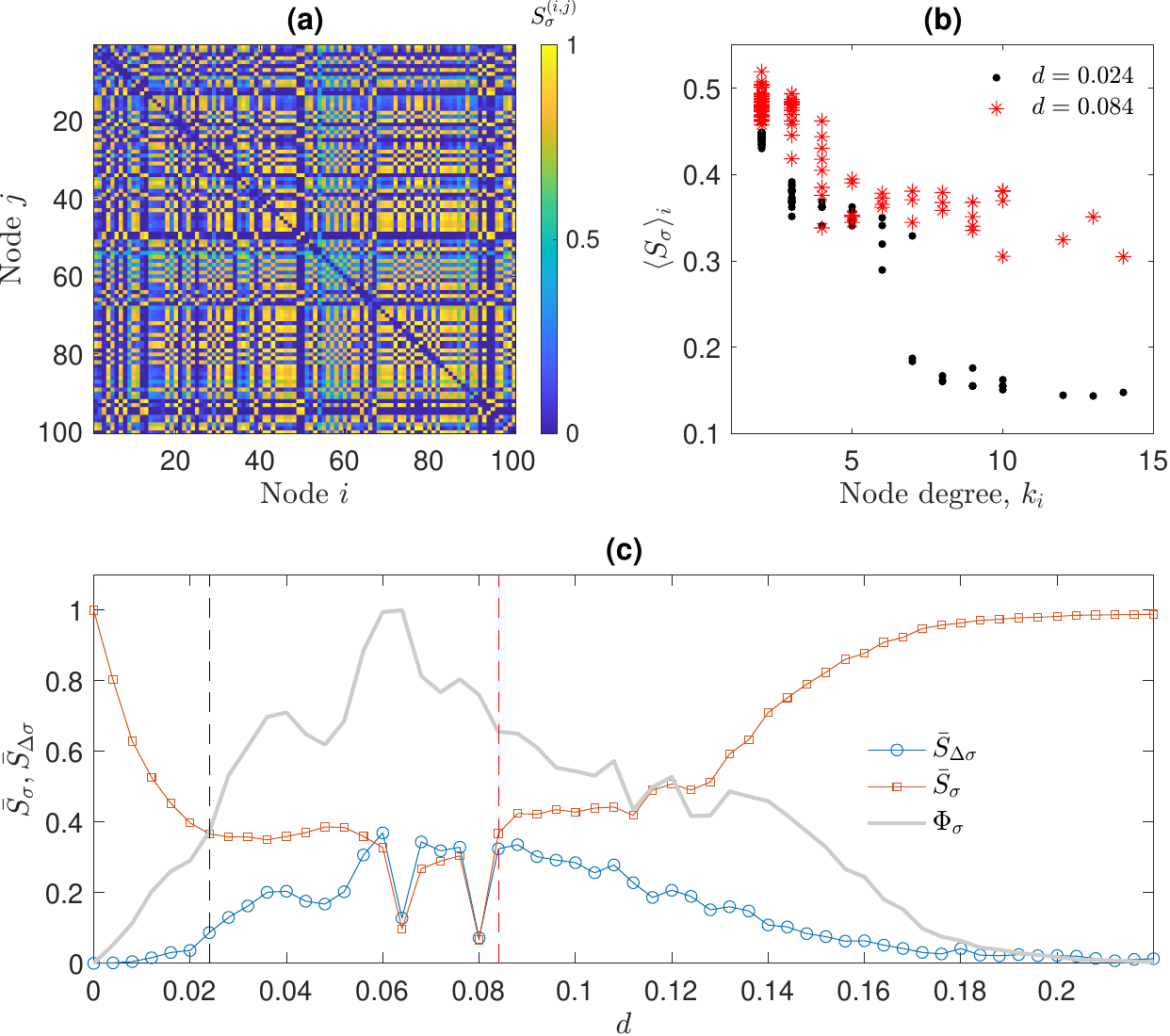}    
  \caption{ (a) Functional symbolic synchronisation matrix $S_{\sigma}^{(i,j)}$ for the same network ensemble used in Fig.~\ref{fig:joint_num_SFER} for $d=0.024$. 
(b) Average functional symbolic synchronisation of each node with its neighbours, $\langle S_{\sigma}\rangle_i$, as a function of their degree $k_i$ for two coupling values shown as vertical lines in panel (c). 
(c) Global averages of symbolic state synchronization, $\bar{S}_{\sigma}$, and jump state synchronization, $\bar{S}_{\Delta\sigma}$, versus coupling $d$. Chaotic itinerancy index $\Phi_{\sigma}$ is plotted in gray for reference. 
  \label{fig:func-sym-sync}}
\end{figure}

The symbolic representation introduced above is not limited to quantifying CI, but it provides a new tool to uncover patterns of coordination that remain hidden to other approaches based on instantaneous phase or amplitude correlations. This perspective allows us to define a  \emph{symbolic synchronization}, where nodes coordinate their dynamics even in the absence or other forms of
synchrony.  We thus examine the extent to which nodes undergo coordinated symbolic transitions, and how it correlates with their structural roles. 

Figure~\ref{fig:func-sym-sync}(a) shows the \emph{functional symbolic synchronization matrix}, $S_{\sigma}$, for the same ensemble of oscillators analyzed in Sec. \ref{seqonset} computed at a representative coupling within the chaotic itinerancy regime ($d = 0.024$). Each matrix element is the normalized Hamming correlation of the respective symbolic sequences: 
\begin{equation}
S^{(i,j)}_{\sigma} = \frac{1}{N_\tau} 
\sum_{\tau=1}^{N_\tau} 
\delta_{\sigma^{(i)}_\tau, \sigma^{(j)}_\tau} \, .
\label{eq:symb-sync}
\end{equation}
High $S^{(i,j)}_{\sigma}$ values indicate frequent symbolic coordination, even in the absence of direct signal synchronization. 

The resulting patterns reveal that nodal symbolic synchronization can emerge 
well before global phase locking, uncovering latent forms of coordination 
driven by the network that can offer information over the underlying topology, 
even for very weak couplings. Figure~\ref{fig:func-sym-sync}(b) displays the 
average symbolic synchronization of each node with its neighbors, 
$\langle S_{\sigma} \rangle_i = N^{-1}\sum_j S^{(i,j)}_{\sigma}$, as a 
function of its degree $k_i$ for two representative coupling strengths [marked 
by vertical lines in Fig.~\ref{fig:func-sym-sync}(c)].  The observed monotonic decrease of $\langle S_{\sigma} \rangle_i$ with $k_i$ indicates that hubs 
synchronize less their regimes than low-degree nodes, which is a 
reflection of the degree-dependent range of symbol exploration 
[Fig.~\ref{fig:symbol_degree_evol}(b)]. 

However, as we have shown before, we can describe another possible level of coordination in the weakly coupled regime: the \emph{itinerancy synchronization}, where nodes change their symbols simultaneously. 
To disentangle this process from the symbolic synchronization we also compute
\begin{equation}
S^{(i,j)}_{\Delta\sigma} =
\frac{1}{N_\tau - 1}\sum_{\tau=1}^{N_\tau-1} 
\delta_{\Delta\sigma^{(i)}_\tau,1}\,\delta_{\Delta\sigma^{(j)}_\tau,1},\label{eq:iti-sync}
\end{equation}
where $\Delta\sigma^{(i)}_\tau = |\sigma^{(i)}_{\tau+1}-\sigma^{(i)}_\tau|$ 
that indicates whether node $i$ presents symbol switched between consecutive 
windows. 
This quantity captures whether nodes perform synchronous symbol switches, 
regardless of which state they occupy before or after the transition.
To compare the two types of synchronization, Fig.~\ref{fig:func-sym-sync}(c) 
shows the evolution as a function of the coupling of the mean values of the 
matrices given by Eqs.\eqref{eq:symb-sync} and \eqref{eq:iti-sync}, 
$\bar{S}_{\sigma}$ and $\bar{S}_{\Delta\sigma}$ respectively. 
In the CI regime ($\Phi_{\sigma}$ is also plotted as a reference), 
$\bar{ S}_{\Delta\sigma}$ rises sharply, showing that transitions become 
temporally aligned across the network, even though $\bar{ S}_{\sigma}$ stays 
lower because nodes often realize different symbols. 
At the end of CI, $\bar{S}_{\sigma}$ becomes high again, as all nodes converge 
to the same symbol, while $\bar{S}_{\Delta\sigma}$ returns to zero since the 
transitions end. It provides a functional viewpoint on how coherence develops 
in complex networks as a process with several stages: first, nodes synchronize 
the \emph{timing} of regime switches (transition synchrony), and later they 
stabilize into the same symbols (state synchrony). 

\section{Experimental evidence}

To experimentally test our predictions, we built a hybrid electronic network composed of $N=28$ analog oscillators emulating R\"ossler-like chaotic dynamics.
The oscillators are interconnected through resistive links that enable adjustable diffusive coupling, with the effective coupling strength \(d\) controlled externally by a programmable voltage source. 
All signal acquisition and feedback operations are handled by a real-time control platform (National Instruments CompactRIO) equipped with high-resolution data acquisition, ensuring precise monitoring and tuning of the local and collective dynamics.

\begin{figure}
    \includegraphics[width=\linewidth]{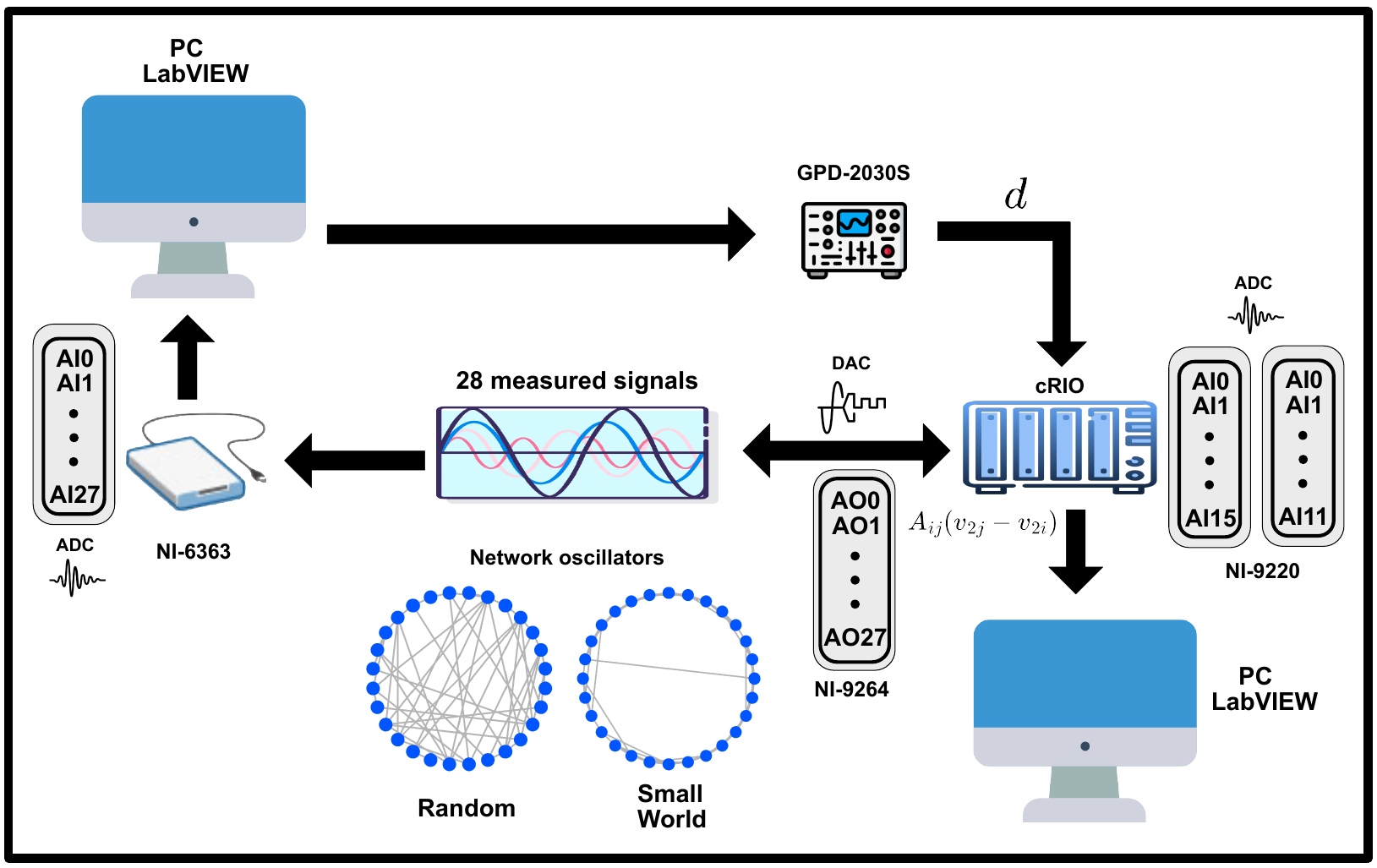} 
  \caption{Schematic of the hybrid experimental acquisition and feedback system. Electronic R\"ossler-like oscillators are interfaced to both a CompactRIO (cRIO-9074) embedded platform and a DAQ board (NI-6363) through their voltage outputs. The cRIO executes the real-time processing required to compute the diffusive coupling terms from the instantaneous node states and generates the feedback signals according to a prescribed adjacency matrix, while the global coupling strength \(d\) is adjusted via an analog control voltage supplied by a programmable power source (GPD-2030S). Signals are recorded through the DAQ for offline analysis and the whole setup is automated with LabVIEW. 
  The hybrid (cRIO+DAQ) architecture enables precise, repeatable sweeps of \(d\) and flexible real-time implementation of different topologies. 
  \label{fig:exp-setup}
   }
    \end{figure}

This hybrid architecture allows precise and repeatable control of the coupling strength, as well as arbitrary reconfiguration of the network connectivity. 
In the present study, we used $N=28$ networks generated according to the Watts–Strogatz model \cite{Watts98} with an average degree $\langle k\rangle = 4$ and two levels of randomness: 
$p = 1$, corresponding to an Erd\H{o}s–R\'enyi–like topology, and $p = 0.1$, which preserves some clustering while remaining largely random. 
Figure~\ref{fig:exp-setup} summarizes the experimental platform and real-time control scheme, while additional details are provided in the Supplementary Information.

\begin{figure}
    \includegraphics[width=\linewidth]{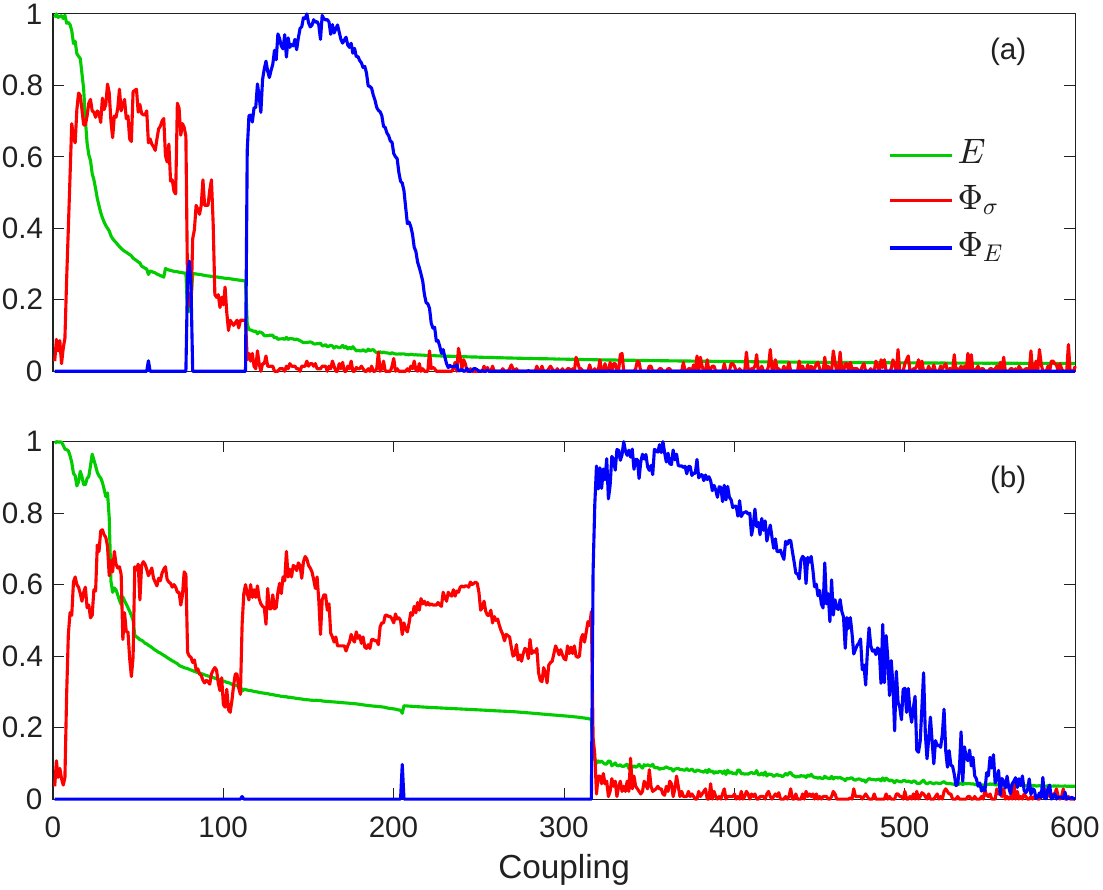} 
  \caption{Normalized mean synchronization error $\langle E \rangle$ (green), chaotic itinerancy index $\Phi_\beta$ (red) and synchronization index $\Phi_E$ (blue) in the experimental $N$=28, $\langle k \rangle$=4 networks with two different topologies: (a) Erd\"os-Renyi, (b) Watts-Strogatz with rewiring probability $p$=0.1. Analysis parameters: $N_\beta=6$, $D=100$, $\theta_{E}=0.02$. }
   \label{fig:joint_exp}
    \end{figure}

Figure \ref{fig:joint_exp} presents the experimental measurements of the chaotic itinerancy index \(\Phi_\sigma\), the intermittent synchronization index \(\Phi_E\), and the average synchronization error \(E\) as a function of the applied coupling for the two implemented topologies. Although the experimental error \(E\) remains finite due to noise and component mismatch, both indices display a clear and robust crossover: at low coupling \(\Phi_\sigma>\Phi_E\) (itinerant-dominated dynamics), while at stronger coupling \(\Phi_E>\Phi_\sigma\) (intermittency-dominated dynamics). 
These observations confirm that the sequence of chaotic itinerancy and intermittent synchronization is not limited to numerical models but constitutes a robust and reproducible feature of real-world weakly coupled chaotic networks.

\section{Conclusions}

In this work, we have demonstrated that the route to synchronization in 
complex networks of chaotic oscillators proceeds through two successive and 
distinct dynamical 
phenomena: \textit{chaotic itinerancy} and 
\textit{intermittent synchronization}. 
Although both phenomena have been previously reported, they are often treated as manifestations of the same underlying process. 
Our results clarify instead that CI and IS constitute \textit{mutually exclusive stages}, each governed by distinct dynamical mechanisms and critical properties, and occurring in separate intervals of the coupling strength.

By introducing a symbolic dynamics, we provided a nodal characterization of 
CI as a collective yet unsynchronized exploration of metastable regimes that 
constitute each nodal sub-attractor. CI, which had not been systematically 
studied in heterogeneous networks, is shown to depend strongly on structural 
diversity: networks with broader degree distributions exhibit richer itinerant 
dynamics, and high-degree nodes have access to larger set of regimes 
than low-degree nodes. A symbolic functional synchronization analysis further 
shows that, during CI, nodes tend to synchronize their \textit{transitions} 
before their \textit{regimes}, highlighting a form of temporal coordination 
that precedes global coherence. The subsequent alignment of symbolic states 
marks the onset of IS before full synchronization.

As coupling increases, CI vanishes and gives rise to IS, where the network 
alternates irregularly between synchronous and desynchronous phases, displaying on--off intermittency. Both regimes exhibit power-law residence times, indicating that critical organization structures the entire synchronization route. The transition between these two phenomena occurs within a consistent coupling 
interval across network topologies and oscillator types, suggesting a robust 
and possibly general scenario. 
An analysis of the Lyapunov spectrum across this transition establishes a 
direct correspondence between the two phenomena and the hierarchical reduction 
of dimensionality that accompanies the emergence of collective order. The experimental implementation in electronic oscillator networks reproduces these features, confirming that the two-step sequence CI--IS is not only a theoretical construct but a generic route observed in real systems.

Beyond its theoretical relevance, the sequential organization of critical 
regimes identified here could be central to understanding flexible 
coordination in biological and artificial systems. 
In the brain, chaotic itinerancy and intermittent synchronization have been 
proposed as mechanisms for cognitive flexibility and information routing 
\cite{Breakspear2017,Calvo2024}. 
Similarly, in reservoir computing and neuromorphic architectures, operating 
near the transition between chaotic itinerancy and intermittent 
synchronization may optimize computational capacity and adaptability 
\cite{Inoue2020,Kong2024,OHagan2025}. 

Overall, our results provide a unified framework linking transient, 
metastable, and critical dynamics in complex networks and highlight the 
fundamental role of structural heterogeneity in shaping the route to 
synchronization.

\bibliography{SysDyn}

\end{document}